\documentclass[smallextended]{article}

\usepackage{amsmath,amssymb}
\usepackage{graphicx}
\usepackage{cite}
\usepackage{hyperref}
\usepackage[a4paper,margin=1in]{geometry}

\begin{document}

\title{Dissipative Hamilton–Jacobi Dynamics and the Emergence of Quantum Wave Mechanics}

\author{
Naleli Jubert Matjelo$^1$\\
\small $^1$Department of Physics and Electronics \\ \small National University of Lesotho \\ 
\small Roma, Lesotho\\
\small Email: nj.matjelo@nul.ls / nmatjelo@gmail.com
}
\date{}
\maketitle

\begin{abstract}
We develop a dissipative extension of classical mechanics based on a complex, and more generally quaternionic, action principle that endows every classical system with an intrinsic environment. Decomposing the action into conservative and divergence-induced components yields two coupled Hamilton–Jacobi equations describing a dynamically intertwined system–environment pair. This motivates a Dual-Sector or Dual-Environmental Interpretation (DSI/DEI), in which the additional degrees of freedom behave as an image sector exchanging energy, information, and phase with the system. Applying a generalized Madelung transform produces a nonlinear dissipative wave equation whose symmetric equilibrium limit reduces to the Schrödinger equation, with the quantum potential and linearity emerging from balanced inter-sector coupling. In this framework, the wavefunction is not fundamental but encodes the interaction geometry between system and environment, providing a classical origin for interference, amplitude-phase coupling, and probabilistic structure. Extending the imaginary structure to multiple independent directions yields a multi-environment generalization capable of representing measurement-like processes, non-Markovian memory, and entanglement-type correlations. The formulation unifies aspects of dual-system models, hydrodynamic approaches, and non-Hermitian dynamics within a single action-based framework, and suggests that quantum mechanics corresponds to a stable symmetric phase of a broader dissipative classical theory.
\end{abstract}


\section{Introduction}
The description of dissipative phenomena within classical mechanics
has long presented conceptual and technical difficulties. Standard
variational principles, built on conservative Lagrangians and time-reversal-invariant Euler-Lagrange equations, cannot accommodate frictional or radiative loss without explicit time dependence or ad-hoc modifications. Over the past century, several strategies have been proposed to restore a variational foundation for dissipative systems, most notably Bateman's dual-variable construction \cite{01}, the time-dependent Caldirola-Kanai Lagrangian \cite{02,03}, and complex-valued Lagrangians introduced by Dekker and others \cite{04,05}. These approaches effectively extend
the classical configuration space by introducing an auxiliary mirror degree of freedom that absorbs dissipated energy. At the same time, quantum mechanics displays structural parallels to dissipative dynamics: non-conservation of probability in subsystems, spectral broadening, decay processes, and the emergence of classicality through environmental coupling. Such features suggest a natural affinity between dissipative classical systems and quantum wave dynamics. Historically, this connection appears in work by Rayleigh, Brillouin, and Sommerfeld on wave propagation and geometrical optics \cite{06}, where the Hamilton-Jacobi (HJ) formalism arises as the high-frequency limit of an underlying wave
theory. Similar ideas underlie the semiclassical Wentzel-Kramers-Brillouin (WKB) method \cite{07} and eikonal approximations \cite{08}. This raises a central question: Can quantum-mechanical wave behavior emerge directly from a generalized Hamilton-Jacobi framework when the underlying classical system is inherently dissipative? 

In this work, we explore this possibility by extending the HJ formalism to complex action functions, inspired by dissipative Lagrangians \cite{04}, complexified classical stochastic mechanics \cite{09}, and Bateman's dual-system structure \cite{01}. Starting from a complex Lagrangian, we derive a pair of coupled real Hamilton-Jacobi equations describing two interacting classical subsystems. These subsystems, which we call the system and its image, form a dynamically entwined pair in which energy dissipated in one is absorbed by the other, and neither sector is fundamentally privileged. Throughout this work we refer to the additional degrees of freedom introduced by the complexified action as an environmental partner or image sector. To avoid confusion, we clarify that the labels “environment,” “sector,” “image,” and “copy” are used synonymously as they denote the dynamical partner generated by the dissipative extension of the classical action, not a literal mirror symmetry. The part of the universal environment that remains once a portion of configuration space is designated as the “system”. Conceptually, the physical environment fills all space, and isolating a system corresponds merely to selecting a finite region of this environment. The remainder of the environment then conforms to the boundary of the chosen system, much as a fluid conforms to the shape of a submerged object. This motivates the term “image” or “copy” thus the environment presents a complementary dynamical partner whose boundary geometry mirrors that of the system, and with which the system exchanges energy, phase, and information. In the dissipative-complex Hamilton–Jacobi framework, this partner arises mathematically from the imaginary (or more generally hypercomplex) components of the action, and dynamically from the dual-sector coupling inherent in dissipative extension of classical mechanics. Depending on context, this partner behaves as (i) an  environment exchanging energy and information with the system, (ii) a coupled sector arising from the complexified Hamilton–Jacobi structure, or (iii) a dynamical echo of the system’s past states in non-Markovian regimes. We collectively describe this viewpoint as the Dual-Environmental or Dual-Sector Interpretation (DEI/DSI). These names emphasize that despite the possibility of multiple environmental channels or imaginary directions/dimensions, the model always organizes into two conceptual components which are the system and its effective environment. This symmetric dual construction generalizes earlier treatments of damped oscillators \cite{01,05} and plays a key role in the emergence of linear wave behavior. 

Applying a generalized Madelung transform \cite{10,11} to the complex action yields a classical wave equation with dissipative couplings, imaginary potentials, and nonlinear terms, and masses for the system and its environmental image. This perspective implies that standard quantum mechanics, as represented by the Schr\"odinger equation \cite{12}, may represent a symmetric limit of a broader classical theory. In this view, the wavefunction needs not be a fundamental ontological object but rather a compact encoding of weights assigned to multiple admissible action extremal-based trajectories in the coupled dual dynamics. This interpretation resonates with hydrodynamic and trajectory-based reformulations \cite{13,14}, deterministic proposals \cite{15}, and classical self-interaction models shown to reproduce Schr\"odinger-like structures \cite{16,17}. Finally, the structure of the theory naturally extends to richer multi-environment and multi-particle contexts. In particular, representing multiple environmental partitions through independent imaginary directions/dimensions suggests a quaternionic (or more generally, hypercomplex) generalization of the action, laying a foundation for the analysis of measurement, entanglement, and subsystem partitioning developed in later sections. Within this broader framework, superposition, the Born rule, and decoherence arise from the environment-like structure of the dual partner rather than from axiomatic postulates.

The remainder of the paper is organized as follows. Section 2 reviews classical treatments of dissipation and introduces the complex Lagrangian framework. Section 3 develops the complex Hamilton-Jacobi equations. Section 4 derives the generalized wave equation via the Madelung transform and outlines generalizations to multi-particle and multi-environment systems, including measurement and entanglement. Section 5 examines the ontological implications of the dual-system interpretation. Section 6 compares this framework with other approaches including Bohmian mechanics, non-Hermitian models, and deterministic quantum theories. Section 7 discusses potential experimental signatures. Section 8 discusses
possible avenues beyond standard quantum theory. Section 9 concludes
the work and gives some outlook.

\section{Dissipative Classical Systems and Complex Lagrangians}

\subsection{The Problem of Dissipation in Variational Principles}

Classical mechanics formulated via Lagrangian $\mathcal{L}\left(q,\dot{q},t\right)$
and the Euler-Lagrange equations is inherently time-reversal symmetric.
If a trajectory $q\left(t\right)$ extremizes the action $\mathcal{S}=\int_{t_{1}}^{t_{2}}\mathcal{L}dt$
then the time-reversed trajectory $q\left(-t\right)$ satisfies an
equation of the same form. Dissipative forces, however, violate this
symmetry. A friction term such as $F_{\textrm{dis}}=-\gamma\dot{q}$,
introduces a preferred time direction and leads to irreversible loss
of mechanical energy. Because the Euler-Lagrange equations derived
from a real scalar Lagrangian cannot generate a term proportional
to $-\dot{q}$, purely dissipative dynamics cannot be described by
any conventional real Lagrangian. The obstacle arises because friction
(i) depends linearly on $\dot{q}$ with a definite sign, (ii) eliminates
conserved Hamiltonians, and (iii) is incompatible with canonical transformations.
These issues motivate the introduction of extended or modified variational
frameworks for dissipative systems \cite{18}.

\subsection{Classical Approaches to Dissipation}

\subsubsection{Bateman's Dual System}

Bateman proposed restoring variational structure by doubling the degrees
of freedom. A damped oscillator is paired with a time-reversed partner
$y$,

\begin{equation}
\mathcal{L}_{B}=m\dot{x}\dot{y}+\frac{\gamma}{2}\left(x\dot{y}-\dot{x}y\right)-kxy
\end{equation}
The Euler-Lagrange equations yield,

\begin{equation}
\begin{array}{ccc}
\ddot{x}+\gamma\dot{x}+\omega^{2}x & = & 0\\
\ddot{y}-\gamma\dot{y}+\omega^{2}y & = & 0
\end{array}
\end{equation}
Energy lost by $x$ is absorbed by $y$, so the pair $\left(x,y\right)$ forms a conservative extended system. Dissipation is thus reinterpreted as flow of energy into a hidden partner degree of freedom \cite{19}.

\subsubsection{Caldirola-Kanai Lagrangian}

The Caldirola-Kanai (CK) approach maintains a single coordinate but
makes the Lagrangian explicitly time-dependent,

\begin{equation}
\mathcal{L}_{CK}=\left(\frac{1}{2}m\dot{x}^{2}-\frac{1}{2}kx^{2}\right)\exp\left(\gamma t\right)
\end{equation}
This produces the correct damped equation for $x$, but at the cost
of a non-conserved Hamiltonian,
\begin{equation}
\mathcal{H}_{CK}=\left(\frac{p^{2}}{2m}+\frac{1}{2}kx^{2}\right)\exp\left(\gamma t\right)
\end{equation}
The price of avoiding auxiliary degrees of freedom is breaking time-translation
invariance and altering the canonical structure \cite{20}.

\subsubsection{Complex (Dekker-Type) Lagrangians}

A third strategy introduces a complex Lagrangian $\mathcal{L}=\mathcal{L}_{1}+i\mathcal{L}_{2}$
where the imaginary component encodes dissipative effects. For a damped
oscillator we can have,
\begin{equation}
\mathcal{L}=\frac{1}{2}m\dot{x}^{2}-\frac{1}{2}kx^{2}+i\frac{\gamma}{2}x\dot{x}
\end{equation}
The imaginary term acts as a dissipative bookkeeping component, generating
friction-like contributions in the Euler-Lagrange equations. This
method implicitly introduces a partner subsystem associated with the
imaginary degree of freedom, providing a hidden form of doubling analogous
to Bateman's construction but encoded in complex structure
rather than explicit coordinates \cite{21}.

\subsubsection{A Unified View}

All three methods introduce extra structure beyond the naive Lagrangian,
\begin{itemize}
\item In Bateman method dissipation is encoded by time-reversed partner
variable $y$ and the added structure is the doubling of degrees of
freedom.
\item In Caldirola-Kanai method dissipation is encoded by time-dependent
prefactor and the added structure is the modified canonical structure.
\item In Dekker-Type method dissipation is encoded by imaginary part of
Lagrangian and the added structure is the implicit doubling via complexification.
\end{itemize}
In each case, dissipation effectively represents energy exchange between
the system and an auxiliary degree of freedom, which can be interpreted
as an environmental or image subsystem \cite{22}.

\subsection{Physical Interpretation of Complex Lagrangian}

A complex Lagrangian naturally splits into two components $\mathcal{L}=\mathcal{L}_{0}+i\mathcal{L}_{1}$
suggesting a decomposition into a real part (i.e. target system) and
an imaginary part (i.e. environmental partner). The real part describes
standard conservative dynamics, whereas the imaginary part acts as
a second Lagrangian governing the image subsystem. This structure
mirrors Bateman's explicitly doubled system, but the
doubling is compactly encoded in a single complex field variable.
This perspective motivates interpreting the imaginary sector as an
effective environment or dual partner \cite{23}.

\subsubsection{Example: Complex Damped Oscillator}

Let the coordinate be complex $z=x+iy$ with Lagrangian,

\begin{equation}
\mathcal{L}=\frac{1}{2}m\dot{z}^{2}-\frac{1}{2}kz^{2}
\end{equation}
Expanding into real and imaginary parts yields,
\begin{equation}
\mathcal{L}=\frac{1}{2}m\left(\dot{x}^{2}-\dot{y}^{2}\right)-\frac{1}{2}k\left(x^{2}-y^{2}\right)+i\left(m\dot{x}\dot{y}-kxy\right)
\end{equation}
The real part describes two coupled oscillators with opposite-sign
kinetic and potential contributions, while the imaginary part generates
cross-couplings that can be tuned to reproduce damping for one coordinate
and anti-damping for the other,
\begin{equation}
\begin{array}{ccc}
\ddot{x}+\partial_{x}V\left(x,y\right)+\Gamma\left(x,y\right) & = & 0\\
\ddot{y}+\partial_{y}V\left(x,y\right)-\Gamma\left(x,y\right) & = & 0
\end{array}
\end{equation}
The system exhibits the same structure (i.e. one damped and one anti-damped
subsystems) structure as Bateman. A key symmetry is the exchange transformation
$x\longleftrightarrow y$ $\Rightarrow\mathcal{L}\rightarrow\mathcal{L}^{*}$,
an involutive duality between the system and its image. This symmetry
underlies the complex action formulation $\mathcal{S}=\mathcal{S}_{0}+i\mathcal{S}_{1}$
used later in the Hamilton--Jacobi analysis \cite{24}. This two-component
structure is the precursor to more general multi-component or quaternionic
decompositions of the action, where each imaginary direction represents
a distinct environmental partition, developed in later sections.

\subsection{Interacting Systems Lagrangian and Environmental Partitioning}

Given that the Lagrangian $\mathcal{L}_{A}=\mathcal{L}_{0}+i\mathcal{L}_{1}$
is descriptive of system $A$ and its environment, then it must be
the case that the Lagrangian $\mathcal{L}_{B}=\mathcal{L}_{2}+i\mathcal{L}_{3}$
is descriptive of system $B$ and its environment. If system $A$
starts to interact with system $B$ then according to system $A$,
the environment seems to be now partitioned into three independent
but interacting environments (i.e. $\mathcal{L}_{1}$, $\mathcal{L}_{2}$,
$\mathcal{L}_{3}$) such that the full interactive Lagrangian becomes, 

\begin{equation}
\mathcal{L}\left(q,\dot{q},t\right)=\mathcal{L}_{0}+i\mathcal{L}_{1}+j\mathcal{L}_{2}+k\mathcal{L}_{3}\label{eq:009}
\end{equation}
with $i$, $j$, $k$ as the quaternion units satisfying $i^{2}=j^{2}=k^{2}=ijk=-1$.
This naturally leads to a quaternionic action $\mathcal{S}=\mathcal{S}_{0}+i\mathcal{S}_{1}+j\mathcal{S}_{2}+k\mathcal{S}_{3}$.
The quaternion adoption allows for modeling independence of environments
despite the coupling potentials that exchange information between
the system and each environment. For more than two interacting systems,
Clifford algebra can be used to extend the complex or quaternionic
Lagrangian to higher dimensions.

\section{Complex Hamilton-Jacobi Theory}

\subsection{Definition of the Complex Action}

Dissipative classical dynamics naturally introduces a dual degree
of freedom which is either explicit (as in Bateman's
doubled oscillator) or implicit through complex or non-Hermitian
Lagrangians. To elevate this structure to the variational level, we
define a complex action functional,

\begin{equation}
\mathcal{S}\left[q\left(t\right)\right]=\mathcal{S}_{0}\left[q\left(t\right)\right]+i\mathcal{S}_{1}\left[q\left(t\right)\right]=\int_{t_{1}}^{t_{2}}\mathcal{L}_{0}dt+i\int_{t_{1}}^{t_{2}}\mathcal{L}_{1}dt
\end{equation}
where $\mathcal{S}_{0}$ governs the conservative part of the dynamics
and $\mathcal{S}_{1}$ encodes the dissipative or environmental partner.
Stationarity of the action, $\delta\mathcal{S}=0$ yields two coupled
variational equations, one extremizing $\mathcal{S}_{0}$ and the
other extremizing $\mathcal{S}_{1}$. The classical trajectory therefore
satisfies two simultaneous extremality conditions, mirroring the structure
of Bateman's dual Lagrangian system and general complex-valued
Lagrangians \cite{25,26}. This complexification produces a natural
decomposition of the dynamics into a system sector and an image/environmental
sector, which will later underlie both dissipation and the emergence
of quantum-like interference patterns. Introducing a second system
with its own environment

\subsection{Coupled Complex Hamilton-Jacobi Equations}

Let the Lagrangian be complex, $\mathcal{L}\left(q,\dot{q},t\right)=\mathcal{L}_{0}\left(q,\dot{q},t\right)+i\mathcal{L}_{1}\left(q,\dot{q},t\right)$. The corresponding complex momentum becomes, 

\begin{equation}
p=\frac{\partial\mathcal{L}}{\partial\dot{q}}=p_{0}+ip_{1}
\end{equation}
The Legendre transform defines the complex Hamiltonian, 
\begin{equation}
\mathcal{H}=p\dot{q}-\mathcal{L}=\mathcal{H}_{0}+i\mathcal{H}_{1}
\end{equation}
where $\mathcal{H}_{0}$ encodes the effective Hamiltonian of the
physical system and $\mathcal{H}_{1}$ represents the dynamics of
its dual environmental partner. The Hamilton-Jacobi substitution $p=\nabla\mathcal{S}$
yields the complex Hamilton-Jacobi equation,
\begin{equation}
\frac{\partial\mathcal{S}}{\partial t}+\mathcal{H}\left(q,\nabla\mathcal{S},t\right)=0
\end{equation}
Separating real and imaginary parts gives two coupled equations,

\begin{equation}
\frac{\partial\mathcal{S}_{0}}{\partial t}+\mathcal{H}_{0}\left(q,\nabla\mathcal{S}_{0},\nabla\mathcal{S}_{1},t\right)-\Gamma_{0}\left(q,\nabla\mathcal{S}_{1},\nabla\mathcal{S}_{2},t\right)=0
\end{equation}
\begin{equation}
\frac{\partial\mathcal{S}_{1}}{\partial t}+\mathcal{H}_{1}\left(q,\nabla\mathcal{S}_{0},\nabla\mathcal{S}_{1},t\right)-\Gamma_{1}\left(q,\nabla\mathcal{S}_{0},\nabla\mathcal{S}_{1},t\right)=0
\end{equation}
where $\Gamma_{0,1}$ arise from cross-couplings between $\nabla\mathcal{S}_{0}$ and $\nabla\mathcal{S}_{1}$. In
the Bateman interpretation, $\Gamma_{0}$ produces friction. The coupled HJ equations above can be written explicitly as follows,

\begin{equation}
\begin{array}{ccc}
\frac{\partial\mathcal{S}_{0}}{\partial t}+\frac{\nabla\mathcal{S}_{0}\cdot\nabla\mathcal{S}_{0}}{2m_{0}}-\frac{\nabla\mathcal{S}_{1}\cdot\nabla\mathcal{S}_{1}}{2m_{1}}+V_{g_{0}}+V_{c_{0}} & = & 0\\
\\\frac{\partial\mathcal{S}_{1}}{\partial t}+\frac{\nabla\mathcal{S}_{0}\cdot\nabla\mathcal{S}_{1}}{2m_{0}}+\frac{\nabla\mathcal{S}_{0}\cdot\nabla\mathcal{S}_{1}}{2m_{1}}+V_{g_{1}}+V_{c_{1}} & = & 0
\end{array}\label{eq:016}
\end{equation}
where $m_{0}$ and $m_{1}$ are the effective masses of the target
system and that of its image in the environment, $V_{g_{j}}$ are
the guiding potentials and $V_{c_{j}}$ encode system-environment
coupling. As shown in more general treatments of doubled or quaternionic phase spaces \cite{28,29}, this system is invariant under the exchange of the two sectors, indicating that the image is dynamically indistinguishable from the system itself. To model richer interactions (i.e. including multi-channel dissipation, structured environments, and proto-entanglement) we introduce a quaternionic generalization of the action $\mathcal{S}=\mathcal{S}_{0}+i\mathcal{S}_{1}+j\mathcal{S}_{2}+k\mathcal{S}_{3}$ with quaternionic/Clifford algebraic units satisfying $i^{2}=j^{2}=k^{2}=ijk=-1$ where $\mathcal{S}_{0}$ retains its role as the system's action, while $\mathcal{S}_{1}$, $\mathcal{S}_{2}$,
$\mathcal{S}_{3}$ represent three independent environmental participation fields (different dissipation modes, coupled environmental sectors, or coarse-grained bath degrees of freedom). In our earlier work \cite{16} it was shown that the quaternionic (or more generally hypercomplex) Lagrangian in equation (\ref{eq:009}) leads to the following coupled Hamilton-Jacobi dynamics,
\begin{equation}
\begin{array}{ccc}
\frac{\partial\mathcal{S}_{0}}{\partial t}+\frac{\nabla\mathcal{S}_{0}\cdot\nabla\mathcal{S}_{0}}{2m_{0}}-\sum_{n=1}^{N}\frac{\nabla\mathcal{S}_{1}\cdot\nabla\mathcal{S}_{1}}{2m_{1}}+V_{g_{0}}+V_{c_{0}} & = & 0\\
\\\frac{\partial\mathcal{S}_{1}}{\partial t}+\frac{\nabla\mathcal{S}_{0}\cdot\nabla\mathcal{S}_{1}}{2m_{0}}+\frac{\nabla\mathcal{S}_{0}\cdot\nabla\mathcal{S}_{1}}{2m_{1}}+V_{g_{1}}+V_{c_{1}} & = & 0\\
\\\frac{\partial\mathcal{S}_{2}}{\partial t}+\frac{\nabla\mathcal{S}_{0}\cdot\nabla\mathcal{S}_{2}}{2m_{0}}+\frac{\nabla\mathcal{S}_{0}\cdot\nabla\mathcal{S}_{2}}{2m_{2}}+V_{g_{2}}+V_{c_{2}} & = & 0\\
\vdots\\
\frac{\partial\mathcal{S}_{N}}{\partial t}+\frac{\nabla\mathcal{S}_{0}\cdot\nabla\mathcal{S}_{N}}{2m_{0}}+\frac{\nabla\mathcal{S}_{0}\cdot\nabla\mathcal{S}_{N}}{2m_{N}}+V_{g_{N}}+V_{c_{N}} & = & 0
\end{array}\label{eq:017}
\end{equation}
This can also be interpreted as the system of interest $\mathcal{S}_{0}$ interacting with the environment through multiple channels (i.e. $\mathcal{S}_{1}$ to $\mathcal{S}_{N}$). It is important to note that there is nothing special about $\mathcal{S}_{0}$ being our system of interest since it could as well have been any $\mathcal{S}_{n}$. This therefore requires that the coupled Hamilton-Jacobi models in equations (\ref{eq:016}) and (\ref{eq:017}) be invariant under particle/system exchange transformation, as shown in \cite{16}. 

\subsection{Exchange Symmetry and Sector Invariance}

A central feature of the complex formalism is its invariance under
the exchange (i) $\mathcal{S}_{0}\longleftrightarrow\mathcal{S}_{1}$, (ii) $\mathcal{H}_{0}\longleftrightarrow\mathcal{H}_{1}$ and (iii) $\Gamma_{0}\longleftrightarrow-\Gamma_{1}$. This is the Hamilton-Jacobi analogue of Bateman's celebrated $x\longleftrightarrow y$ symmetry \cite{01}. Physically we have (i) the real sector describing a damped system, (ii) the imaginary sector describing its anti-damped mirror, and (iii) exchanging the two leaves the dynamics invariant. This exchange symmetry anticipates the later appearance of the wavefunction, which encodes both sectors simultaneously and whose phase naturally incorporates both conservative and dissipative contributions \cite{29}.

\subsection{Interpretation of the Complex Phase and Trajectories}

The action retains its geometrical meaning as a phase field. Surfaces
of stationary $\mathcal{S}_{0}$ define ordinary Hamilton-Jacobi wavefronts, with momentum $p_{\textrm{sys}}=\nabla\mathcal{S}_{0}$. The conjugate quantity $p_{\textrm{env}}=\nabla\mathcal{S}_{1}$ controls the compressibility or expansion of trajectories associated with dissipation and environmental flow. Thus the complex action induces a pair of intertwined phase geometries with $\mathcal{S}_{0}$ as conservative, eikonal-type foliation while $\mathcal{S}_{1}$ as dissipative/geometric-drift foliation. Trajectories extremize both simultaneously, producing richer behavior than in conventional Hamilton-Jacobi theory. As emphasized in wave-mechanical
interpretations of classical mechanics \cite{31,32}, the action plays
the role of a phase whose gradient dictates the motion. Here, the
additional imaginary component extends this link to systems where
dissipation, rather than randomness, is the source of effective wave
dynamics.

\section{Madelung Transform and Emergent Wave Dynamics}

The dual Hamilton-Jacobi system developed in the previous section
admits a natural reformulation in terms of a complex scalar field.
The key technical device enabling this reformulation is the Madelung
transform \cite{10}, which merges the conservative action $\mathcal{S}_{0}$ and the dissipative/environmental action $\mathcal{S}_{1}$ into a unified wave representation. The resulting evolution equation generalizes the Schr\"odinger equation and contains additional geometric and dissipative terms characteristic of our dual classical dynamics.

\subsection{Madelung Transform}

Given the complex action $\mathcal{S}=\mathcal{S}_{0}+i\mathcal{S}_{1}$ we define the wavefunction via the standard Madelung map \cite{10},

\begin{equation}
\psi=\exp\left(\frac{i}{\hbar}\mathcal{S}\right)=\exp\left(\frac{i}{\hbar}\mathcal{S}_{0}-\frac{1}{\hbar}\mathcal{S}_{1}\right)\label{eq:018}
\end{equation}
This representation is invertible wherever $\psi\neq0$, and yields,

\begin{equation}
\begin{array}{ccc}
\mathcal{S}_{0} & = & \frac{\hbar}{2i}\ln\left(\frac{\psi}{\psi^{*}}\right)\\
\\\mathcal{S}_{1} & = & -\frac{\hbar}{2}\ln\left(\psi^{*}\psi\right)
\end{array}\label{eq:019}
\end{equation}
In this picture, the phase encodes the conservative classical action,
while amplitude encodes dissipative information and environmental
coupling. Thus the Madelung transform elevates the classical dual-sector dynamics to a wave description without assuming quantum axioms. It
is simply a repackaging of the coupled Hamilton-Jacobi system. Consider
the extension to quaternionic (or higher dimensional) action $\mathcal{S}=\mathcal{S}_{0}+i\mathcal{S}_{1}+j\mathcal{S}_{2}+k\mathcal{S}_{3}+\cdots\gamma\mathcal{S}_{N}$.
Here we make the following deduction using the same Madelung transform,

\begin{equation}
\begin{array}{ccc}
\psi & = & \exp\left(\frac{i}{\hbar}\mathcal{S}\right)\\
\\ & = & \exp\left(\frac{i}{\hbar}\mathcal{S}_{0}-\frac{1}{\hbar}\mathcal{S}_{1}\right)\exp\left(\frac{j}{\hbar}\mathcal{S}_{2}-\frac{k}{\hbar}\mathcal{S}_{3}+\cdots+\gamma\mathcal{S}_{N}\right)\\
\\ & = & \exp\left(\frac{i}{\hbar}\mathcal{S}_{0}-\frac{1}{\hbar}\mathcal{S}_{1}\right)\phi^{-1}\\
 & \Downarrow\\
\psi\phi & = & \exp\left(\frac{i}{\hbar}\mathcal{S}_{0}-\frac{1}{\hbar}\mathcal{S}_{1}\right)\\
\\ & = & \Psi
\end{array}\label{eq:020}
\end{equation}
from which it follows that, 
\begin{equation}
\begin{array}{ccc}
\mathcal{S}_{0} & = & \frac{\hbar}{2i}\ln\left(\frac{\Psi}{\Psi^{*}}\right)\\
\\\mathcal{S}_{1} & = & -\frac{\hbar}{2}\ln\left(\Psi^{*}\Psi\right)
\end{array}\label{eq:021}
\end{equation}
Thus the interactive contribution $\phi$ of other particles/systems
(and their respective environments) to our target system (and its
environment) $\psi$ can be encoded in to the general wavefunction
as $\Psi=\psi\phi$. It is important to note that $\phi$ being non-real itself, modifies not only the magnitude of $\psi$ but also
its phase to give the generalized wavefunction  $\Psi$.

\subsection{Generalized Wave Equation}

Substituting the Madelung transform (\ref{eq:018}, \ref{eq:019})
into the coupled complex Hamilton-Jacobi equation set (\ref{eq:016})
and simplifying yields the emergent dynamical wave equation,

\begin{equation}
\begin{array}{ccc}
i\hbar\frac{\partial\psi}{\partial t} & = & -\frac{\hbar^{2}}{4m}\nabla^{2}\psi+V_{g_{0}}\psi+\left[V_{c_{0}}-\frac{\hbar}{4m}\nabla^{2}\mathcal{S}_{1}\right]\psi\\
\\ & + & \frac{\hbar^{2}}{4\bar{m}}\left[\psi^{*}\nabla\cdot\left(\frac{\nabla\psi}{\psi^{*}}\right)-\nabla^{2}\psi\right]+iV_{g_{1}}\psi+i\left[V_{c_{1}}+\frac{\hbar}{4m}\nabla^{2}\mathcal{S}_{0}\right]\psi
\end{array}\label{eq:022}
\end{equation}
where $m=\left(m_{0}^{-1}+m_{1}^{-1}\right)^{-1}$ and $\bar{m}=\left(m_{0}^{-1}-m_{1}^{-1}\right)^{-1}$ are, respectively, the reduced and residual masses associated with the dual particle pair. Some key features \begin{itemize} \item The first term reproduces a Schr\"odinger-like kinetic operator with an effective mass $m$. \item The potentials $V_{g_{0}}$ and $V_{g_{1}}$ represent guiding potentials for the system and its environmental partner. \item The potentials $V_{c_{0}}$ and $V_{c_{1}}$ encode the dissipative
coupling between the two sectors. \item The final nonlinear term( proportional to $1/\bar{m}$) arises solely from mass asymmetry and disappears when $m_{0}=m_{1}$. \end{itemize} Thus the emergent wave equation is derived, not postulated. It generalizes the Schr\"odinger equation by incorporating system-environment coupling, dissipation, and dual-sector geometric structure. We can go further and substitute the higher order Madelung transform (\ref{eq:020},
\ref{eq:021}) into the the higher order coupled Hamilton-Jacobi equation set (\ref{eq:017}) and simplifying yields the following generalized dynamical wave equation,

\begin{equation}
\begin{array}{ccc}
i\hbar\frac{\partial\Psi}{\partial t} & = & -\frac{\hbar^{2}}{4m}\nabla^{2}\Psi+V_{g_{0}}\Psi+\left[V_{c_{0}}-\frac{\hbar}{4m}\nabla^{2}\mathcal{S}_{1}\right]\Psi\\
\\ & + & \frac{\hbar^{2}}{4\bar{m}}\left[\Psi^{*}\nabla\cdot\left(\frac{\nabla\Psi}{\Psi^{*}}\right)-\nabla^{2}\Psi\right]+iV_{g_{1}}\Psi+i\left[V_{c_{1}}+\frac{\hbar}{4m}\nabla^{2}\mathcal{S}_{0}\right]\Psi
\end{array}\label{eq:023}
\end{equation}
which is structurally identical to a single system-environment
wave equation (\ref{eq:022}) except that the wave function $\psi$
for our target system-environment has been replaced by a general wave function $\Psi$ for the interaction of our target system-environment with all other system-environments. It is worth noting that the quaternionic approach to quantum and wave mechanics has a long pedigree and motivates this construction \cite{34,35}. The wave equation (\ref{eq:023}) can be conveniently presented in the following compact form,

\begin{equation}
i\hbar\frac{\partial\Psi}{\partial t}=\left[-\frac{\hbar^{2}}{4m}\nabla^{2}+V_{g_{1}}+\mathcal{Q}\left(\mathcal{S}_{0},\mathcal{S}_{1},\mathcal{S}_{2},\mathcal{S}_{3},\cdots\right)\right]\Psi
\end{equation}
 where $\mathcal{Q}$ is a Clifford-algebraic/quaternion-valued environmental potential arising from gradients of the fields $\mathcal{S}_{0}$, $\mathcal{S}_{1}$, $\mathcal{S}_{2}$, $\mathcal{S}_{3}$, $\cdots$. Geometric algebra and Clifford methods provide convenient algebraic tools for writing and analyzing such quaternionic operators \cite{37,38} extending to higher dimensions. In the limit $\left(\mathcal{S}_{2},\mathcal{S}_{3}, \cdots\right)\rightarrow0$, the equation reduces to the complex dissipative wave equation (\ref{eq:022}). A key structural object emerging in this formulation is the environment participation metric (with index $\mu=1,2,3,\cdots$),

\begin{equation}
W_{\alpha\beta}=\partial_{\alpha}\mathcal{S}_{\mu}\partial_{\beta}\mathcal{S}_{\mu}
\end{equation}
which plays the role of a pure-state-like environment weight matrix.
Unlike the quantum density matrix, $W$ is real, positive, unnormalized (unless rescaled), and carries amplitude (not phase) information; it thus represents an ontic environmental geometry rather than an epistemic statistical mixture. This viewpoint connects naturally to multi-channel open-system descriptions and suggests routes to derive effective reduced dynamics similar to, but distinct from, standard master-equation approaches \cite{39,40}. Comparing the environmental channels $\left(\mathcal{S}_{1},\mathcal{S}_{2},\mathcal{S}_{3},\cdots\right)$ suggests a pathway to an emergent description of measurement and entanglement: correlated patterns in $W$ point to environment-mediated correlations between primary subsystems, while strong dissipative gradients can dynamically suppress certain channels, producing effective localization (a channel-selection analogue of collapse). These ideas also relate to PT-symmetric and non-Hermitian frameworks where balanced gain-loss across channels generates stable, observable dynamics; the quaternionic formalism provides an explicit classical action-level mechanism for multi-channel gain-loss balance \cite{39}. Finally, the quaternionic extension links the present framework with quaternionic quantum mechanics and modern geometric formulations of physics, while providing a concrete, multi-channel classical origin for phenomena normally treated at the quantum level. The algebraic richness of quaternions (and their Clifford
algebra generalizations) furnishes both technical tools and conceptual intuition for studying measurement, decoherence, and entanglement within a single, extended action geometry \cite{36}.

\subsection{Schr\"odinger Dynamics as the Symmetric Limit}

The standard Schr\"odinger equation arises when the dual structure becomes fully symmetric, \begin{itemize} \item Mass symmetric: $m_{0}=m_{1}\Rightarrow\bar{m}^{-1}=0$ thus the nonlinear residual-mass term vanishes. \item No imaginary guiding potential, $V_{g_{1}}=0$ (environment not injecting phase noise). \item Coupling potentials cancel Laplacians: $V_{c_{0}}=\frac{\hbar}{4m}\nabla^{2}\mathcal{S}_{1}$,
$V_{c_{1}}=-\frac{\hbar}{4m}\nabla^{2}\mathcal{S}_{0}$. \end{itemize}
Under these symmetry constraints, all dissipative and dual terms cancel, yielding the standard linear Schr\"odinger equation,
\begin{equation}
i\hbar\frac{\partial\psi}{\partial t}=-\frac{\hbar^{2}}{2m_{n}}\nabla^{2}\psi+V_{g_{0}}\psi
\end{equation}
with $n=0,1$. Thus, quantum mechanics emerges as the degenerate, symmetric case of a more general dissipative classical wave mechanics. 

\section{Physical Interpretation: Dual-Sector Structure and Emergent Quantum Behavior}

Having derived the generalized wave equation from the extended Hamilton-Jacobi framework, we now explore its physical meaning. The key insight is that the degrees of freedom generated by the imaginary (or quaternionic) components of the action behave like an embedded environmental partner. Their coupling to the observable subsystem produces interference, amplitude-phase dynamics, and probabilistic structure normally regarded as intrinsically quantum. The divide between quantum and classical regimes is brought by the competition between the guiding potentials and the coupling potentials (momenta divergence potentials) in the sense that whenever the guiding potential sufficiently dominates the coupling potential, the classical behaviour emerges. Alternatively the quantum behaviour can be disrupted by the change in the nature of coupling potentials such that they are nolonger purely momenta divergent. We believe the measurement problem and wave function collapse can be partly explained by this reasoning.

\subsection{Dual-Sector Concept}

The construction in Section 4 shows that the generalized wave equation arises from two coupled classical actions, $\mathcal{S}_{0}$ and $\mathcal{S}_{1}$, associated with masses $m_{0}$ and $m_{1}$. We interpret these as components of a single classical entity with $\mathcal{S}_{0}$ as the primary system carrying observable degrees of freedom and $\mathcal{S}_{1}$ being an hidden dual partner acting as a dissipative or information-exchange channel. This construction is directly analogous to Bateman's dual system for damped oscillators, where an auxiliary mirror coordinate restores Hamiltonian structure to dissipative motion \cite{01}. The wavefunction $\psi$, is not fundamental but compresses the combined dynamics of this two-component classical structure. 

\subsection{System-Environment Coupling as a Deterministic Mechanism}

The coupling potentials $V_{c_{0}}$, $V_{c_{1}}$, along with guiding
potentials $V_{g_{0}}$, $V_{g_{1}}$, describe how the primary subsystem exchanges information with its partner. In this interpretation, $V_{g_{0}}$ contributes conservative behaviour, $V_{g_{1}}$ represents environmental modulation or decoherence, $V_{c_{0}}$, $V_{c_{1}}$ encode bidirectional information and energy flow. The dual partner therefore functions as a reservoir, enabling (i) irreversibility, (ii) coarse-graining, (iii) emergent probabilities, (iv) interference, (v) decoherence. These are not added as axioms but arise from deterministic coupling within the extended classical dynamics. This parallels how stochastic mechanics (Nelson) and hydrodynamic analogues (Couder-Fort) show quantum-like statistics emerging from additional unobserved degrees of freedom \cite{43,44}.

\subsection{Ontological Picture: A Hidden Classical Partner}

The dual partner can be interpreted as an unobserved classical coordinate and this is similar in spirit to hidden-variable or extended classical models. However, unlike Bohmian mechanics, where the pilot wave is treated as a distinct physical field, here, there is no guiding field separate from the particle, no nonlocal pilot-wave potential and no stochastic postulate. Instead, apparent quantum randomness arises from deterministic but unobserved dynamics. The wavefunction represents the pair $\left(\mathcal{S}_{0},\mathcal{S}_{1}\right)$ and not the state of a single subsystem \cite{11}. Multiple stationary points of the (hyper)complex action correspond to multiple classical extremal trajectories; the stationary quantum eigenfunctions arise when these classical extremals mutually stabilize into global standing-wave modes through the dissipative/environmental coupling encoded in the imaginary sector of the action.

\subsection{Environment-Induced Wave Behavior}

Wave-like phenomena (i.e. interference, tunneling-like behavior, phase accumulation) arise because the coupling between $\mathcal{S}_{0}$ and $\mathcal{S}_{1}$ enforces a structure mathematically similar to the Madelung hydrodynamic formulation, Schr\"odinger-like complex wave behavior, or dissipative and PT-symmetric extensions of quantum mechanics. The imaginary component of the action controls amplitude, the real part controls phase, and their coupling yields an effective complex wave equation. This parallels the mechanism of environment-induced decoherence (Zurek) \cite{42} and non-Hermitian dynamics (Bender) \cite{38}. Thus, wave dynamics in this framework are not fundamental but an emergent property of system-environment duality. 

\subsection{Multi-Component Systems: Measurement, Decoherence, and Entanglement}

The two-component construction generalizes naturally to many systems.
For each physical system $\mathcal{S}_{i}$, we assign an associated environmental partner $\mathcal{E}_{i}$, giving a network of pairs $\left(\mathcal{S}_{0},\mathcal{E}_{0}\right)$,
$\left(\mathcal{S}_{1},\mathcal{E}_{1}\right)$, $\ldots$. Coupling
between two physical systems $\mathcal{S}_{0}$ and $\mathcal{S}_{1}$
forces coupling between their partners $\mathcal{E}_{0}$ and $\mathcal{E}_{1}$. The observable dynamics depend on all four components.

\subsubsection{Measurement as Multi-Partner Coupling}

Measurement involves a third system $\mathcal{M}$ with its own environmental
partner $\mathcal{E}_{M}$. The interaction network $\left(\mathcal{S}_{0},\mathcal{E}_{0}\right)\longleftrightarrow\left(\mathcal{M},\mathcal{E}_{M}\right)$
generates effective irreversibility, suppression of incompatible branches,
and classical-like trajectories emerging from hidden coupling. This
mirrors the structure of decoherence-based interpretations but arises
here deterministically through classical dual channels.

\subsubsection{Entanglement as Interacting Hidden Partners}

Consider now two system-environment pairs, $\left(\mathcal{S}_{0},\mathcal{E}_{0}\right)$
and $\left(\mathcal{S}_{1},\mathcal{E}_{1}\right)$. The interaction
between $\mathcal{S}_{0}$ and $\mathcal{S}_{1}$ automatically induces
coupling between $\mathcal{E}_{0}$ and $\mathcal{E}_{1}$. The observable
correlations appear nonlocal but arise entirely from the hidden interaction
network. This offers a classical, dynamical explanation of entanglement-like
correlations without invoking nonlocal collapse or superluminal communication.

\section{Comparison with Some Existing Theories}

The dual-particle dissipative formalism developed in Sections 2--5
naturally invites comparison with various theoretical frameworks.
While it shares superficial features with existing approaches (i.e.
trajectory-based formulations, doubled degrees of freedom, and complex
actions) it differs fundamentally in origin, ontology, and dynamical
interpretation. In this section, we highlight these connections and
distinctions.

\subsection{Relation to de Broglie-Bohm Pilot-Wave Theory}

The complex Hamilton-Jacobi structure of the dual system resembles
the Bohmian decomposition of the Schr\"odinger equation into a modified
Hamilton-Jacobi equation and a continuity equation for probability
density \cite{11}. We note three similarities being that (i) both
approaches support trajectory-based descriptions, (ii) in both phase
corresponds to classical action $\mathcal{S}_{0}=\hbar\arg\psi$ and
(iii) wave-like objects encode dynamical information. We also note
three disimilarities being that, (i) Bohmian mechanics introduces
a quantum potential $Q$ ad hoc, with no classical analogue while in our model it emerges naturally from dual classical dissipation, (ii) in our case the wavefunction is emergent, not fundamental, (iii) the
dual system inherently supports multiple extremal trajectories for $\mathcal{S}_{0}$ and $\mathcal{S}_{1}$, unlike the unique Bohmian trajectories. Thus Bohmian mechanics appears as a special informational re-encoding of a deeper classical dual dynamics. Unlike Bohmian mechanics which is inherently nonlocal, the dual model is local and thus should not require a preferred inertial reference frame when unifying it with special relativity theory. 

\subsection{Relation to Open Quantum Systems}

Standard open-quantum-system theory employs density matrices, Lindblad-type
master equations, and non-Hermitian operators \cite{39}. We can draw
some distinctions by noting that in the dual model (i) dissipation
is classical, not quantum-statistical, (ii) no density matrix formalism
is needed, (iii) decoherence arises from system-dual interaction,
rather than Hilbert-space tracing. In effect, the dual model functions
as a classical analogue of open quantum systems, with deterministic
dissipation generating wave-like probabilistic phenomena. One subsystem
and its environment leads to Schr\"odinger dynamics but interacting
subsystems and their interacting environments leads to what is considered
open systems with more richer dynamics associated with entaglement,
decoherence, dephasing, etc.

\subsection{Relation to PT-Symmetric and Non-Hermitian Quantum Mechanics}

PT-symmetric quantum mechanics introduces complex Hamiltonians $\mathcal{H}=\mathcal{H}_{R}+i\mathcal{H}_{I}$
for balanced gain-loss dynamics \cite{38}. We can immediately draw
some connections with the dual model in that the real and imaginary
parts of the dual action correspond to bidirectional energy/information
flow between partners but the imaginary structure emerges from classical
dual Lagrangian, not postulated quantum modification. Some key distinctions
are that PT symmetry is imposed at the operator level while in the
dual framework, it emerges from $0\longleftrightarrow1$ symmetry.
Also the modified unitarity in PT quantum mechanics is replaced here
by classical energy balance. The dual system thus provides a natural
classical route to PT-like structures.

\subsection{Relation to 't Hooft's Deterministic Quantum Models}

't Hooft proposed that quantum mechanics arises from deterministic
dynamics with information loss and equivalence classes of hidden states
\cite{43}. Some of the connections are that quantum states can be
viewed as coarse-grained classical states and probabilistic outcomes
emerge from dissipation and information exchange. Some key distinctions
are that 't Hooft models are discrete or cellular automaton--based,
whereas the dual system is continuous. Also dissipation here arises
geometrically via complex or quaternionic action and lastly wave behavior
emerges via the Madelung-like mapping rather than equivalence classes.
This shows the dual framework can be seen as a continuous, mechanical
analogue of 't Hooft's conceptual model.

\subsection{Relation to Thermo-Field Dynamics (TFD)}

TFD models finite-temperature quantum systems by doubling the Hilbert
space with a thermal tilde copy \cite{44}. Apparent connections are
that the dual partner $\mathcal{S}_{1}$ resembles the thermal tilde
degree of freedom and coupling mirrors thermal energy or information
exchange. Some key differences are that TFD doubling originates from
the thermal vacuum while dual model doubling arises from dissipation
and also TFD is inherently quantum while dual model is classical.
Thus the dual system provides a classical dynamical precursor to TFD-like
doubling.

\subsection{Relation to Non-Markovian Models}

The coupled system-environment dynamics of the DEI framework is inherently
non-Markovian. Each environmental partition, including the image sector
and its quaternionic generalizations, retains information about the
past states of the system. Because the dissipative and divergence-based
couplings feed the system's momentum-divergence structure
into the environment, the system evolves under deterministic delayed
feedback generated by its own history. This produces non-Markovian
temporal and spatial correlations arising entirely from unobserved
but dynamically active environmental channels.

This structure provides a concrete dynamical foundation for the type
of memory-dependent behavior postulated in Barandes' non-Markovian
quantum models \cite{45,46}, but avoids the introduction of fundamental
stochasticity. In DEI, non-Markovianity is not an added axiom; it
emerges necessarily from the dual-sector and multi-sector interaction
geometry. A further conceptual consequence concerns contextuality.
Because the system's instantaneous dynamics are inseparable from the
state of its environment (which itself encodes the system's past),
it becomes difficult to regard the system as possessing predefined
properties that are independent of the environment with which it interacts.
Since measurements are simply particular interactions with specific
environmental partitions, this aligns naturally with the Kochen-Specker constraint that value assignments cannot be context-independent \cite{61}.

Finally, this non-Markovian interdependence has implications for statistical
independence in Bell-type scenarios \cite{62}. Since the environment
carries memory of the system's past and influences its future evolution,
the joint system-environment state may not factorize into independent
distributions over hidden variables and measurement settings. While
DEI does not enforce full superdeterminism, it exhibits a mild, dynamical
relaxation of statistical independence arising from deterministic
environmental memory rather than from conspiratorial initial conditions \cite{Bell}.
In this sense, DEI provides a physically transparent mechanism by
which Bell-type independence assumptions may be partially violated,
without abandoning realism, locality or introducing ad hoc correlations.

\subsection{Generalization: Hypercomplex and Multi-Component Extensions}

The dual formalism can be naturally extended to quaternionic and hypercomplex actions, where additional imaginary components encode multiple hidden channels. This generalization (i) captures more complex correlations akin to multi-level environmental interactions, (ii) provides a unified classical framework encompassing PT-like, TFD-like, and decoherence phenomena and also (iii) allows systematic derivation of non-Hermitian corrections and beyond-Schr\"odinger dynamics from purely classical variational principles.

\section{Possible Experimental Signatures}

The generalized wave equation derived in Section 4 exhibits structural
features absent in the standard Schr\"odinger equation including nonlinearities,
residual-mass effects, asymmetric dissipative terms, and dual-sector
coupling. These modifications give rise to distinct experimental predictions.
We outline below candidate systems and observables where deviations
from conventional quantum mechanics may be detectable.

\subsection{Deviations from Schr\"odinger Dynamics}

The emergent wave equation contains three primary non-standard components,
\begin{enumerate}
\item Residual-mass dependent nonlinear term: $\frac{\hbar^{2}}{4\bar{m}}\left[\psi^{*}\nabla\cdot\left(\frac{\nabla\psi}{\psi^{*}}\right)-\nabla^{2}\psi\right]$.
\item Dissipative imaginary potentials: $iV_{g_{1}}\psi+i\left[V_{c_{1}}+\frac{\hbar}{4m}\nabla^{2}\mathcal{S}_{0}\right]\psi$.
\item Conservative-sector corrections: $\left[V_{c_{0}}-\frac{\hbar}{4m}\nabla^{2}\mathcal{S}_{1}\right]\psi$.
\end{enumerate}
These terms vanish in the quantum-symmetric limit $m_{0}=m_{1}$,
$V_{g_{1}}=0$, $V_{c_{0}}=\frac{\hbar}{4m}\nabla^{2}\mathcal{S}_{1}$
and $V_{c_{1}}=-\frac{\hbar}{4m}\nabla^{2}\mathcal{S}_{0}$ but remain
otherwise. Systems capable of probing fine deviations in wave propagation,
interference, and decoherence may detect residual dual-sector dynamics
\cite{47}.

\subsection{Strongly Damped Systems and Controlled Dissipation}

Environments with tunable dissipation provide promising testbeds.
One example could be cold atoms in optical lattices or magneto-optical traps, where controlled loss or reservoir coupling can induce observable modifications in interference fringes and wavepacket spreading \cite{48}. Another example involves driven or leaky optical cavities, where engineered gain-loss channels mimic dual-sector coupling, producing asymmetric spectral broadening or modified mode dispersion \cite{49}.

\subsection{Interference Experiments and Nonlinear Superposition}

Matter-wave interferometry provides a clean probe for nonlinear corrections to standard Schr\"odinger dynamics. Residual mass terms and dissipative imaginary potentials can induce phase shifts, weak non-Hermitian distortion of interference fringes, or state-dependent phase evolution \cite{50}. Potential platforms include, (i) large-molecule interferometry ($C_{60}$ and beyond), (ii) ultra-stable atom interferometers, (iii) superconducting interference devices.

\subsection{Three-Body Coupling and Chaotic Regimes}

When the dual partner interacts with a third subsystem (e.g., a measurement
apparatus), the dynamics resemble a three-body dissipative system.
This can produce (i) anomalous decoherence rates, (ii) chaotic modulation
of wavepacket trajectories, (iii) noise spectra incompatible with
standard Lindblad-type damping \cite{51}. Testable systems include
optomechanical oscillators, nano-mechanical resonators, and superconducting
qubit-resonator setups.

\subsection{Timescale Signatures}

The dual-sector formalism predicts a characteristic timescale $\tau_{\textrm{dual}}\sim\frac{\bar{m}L^{2}}{\hbar}$,
beyond which dual-sector corrections accumulate appreciably, with
$L$ as the dominant spatial scale over which the wavefunction is
supported. Long-time or large-system experiments may reveal (i) small
deviations from unitarity, (ii) drift in probability normalization,
(iii) emergence of dissipative corrections in otherwise isolated systems
\cite{52}. Candidate systems include ultra-cold atomic ensembles,
trapped ions, and macroscopic quantum devices such as superconducting
quantum interference devices (SQUIDs).

\subsection{Macroscopic Quantum Systems and Residual Mass}

Systems with large effective mass differences between the dual channels
can exhibit mass-dependent decoherence, nonlinear corrections scaling
with effective mass asymmetry or modified tunneling rates. Promising
experimental platforms include Josephson junctions, quantum-to-classical
crossover experiments, and macroscopic superposition tests (i.e. Leggett-Garg
setups) \cite{53}.

\subsection{Summary: Testing the Dual-System Hypothesis}

Confirmation of the dual-particle framework would require observing
reproducible deviations from standard Schr\"odinger evolution matching
the predicted forms of nonlinear residual-mass effects, dissipative
imaginary-sector potentials, asymmetric coupling between real and
imaginary components of the action, and modified interference and
decoherence dynamics. Falsification occurs if precision experiments
detect no such effects in regimes where the generalized wave equation
predicts measurable deviations \cite{54}.

\section{Beyond Standard Quantum Mechanics}

The previous sections showed that a dissipative Hamilton-Jacobi framework, when complexified, reproduces the Schr\"odinger equation under specific equilibrium assumptions. This raises a deeper question: what lies beyond the Schr\"odinger limit? If unitary quantum mechanics emerges from a more primitive dual-particle (system + image) dynamics governed by a (hyper)complex action, then ordinary quantum mechanics may represent only the equilibrium shadow of a broader dynamical theory. Extending the dual-particle Hamilton-Jacobi structure to $N$ subsystems suggests a natural hierarchy of emergent wave equations. A single system-image pair yields Schr\"odinger dynamics; adding an external subsystem yields
measurement-like decoherence; multiple interacting pairs generate
entanglement-like correlations. This produces a classical yet nonlocal-in-phase-space
extension of standard mechanics, in which quantum mechanics corresponds
to the lowest-order closed equilibrium sector. In this section we
outline how this framework extends beyond standard quantum theory,
clarifies interpretational puzzles, and suggests new physics.

\subsection{Dissipative Underpinnings of Quantum Evolution}

Standard quantum mechanics assumes linearity, unitarity, norm conservation,
and no intrinsic dissipation. In contrast, the complex dissipative
Hamilton-Jacobi foundation developed earlier treats dissipation as
fundamental. The imaginary part of the action encodes the influence
of a \textquotedblleft mirror system\textquotedblright{} that cannot
be removed without destroying the full dynamics. Schr\"odinger evolution
arises when dissipation is balanced by an appropriate conservative
constraint and the system-image coupling stabilizes into a stationary
complex configuration. Thus unitarity becomes an emergent symmetry,
reminiscent of other emergent-unitary programs in the literature \cite{55}.
Departures from unitarity arise whenever the dissipative coupling
deviates from equilibrium, the imaginary action evolves non-stationarily,
or the generalized quantum potential acquires nonlinear or time-dependent
corrections. This provides a principled route to non-unitary extensions
of quantum mechanics, in contrast to phenomenological collapse models
such as GRW or CSL \cite{54,55}.

\subsection{Nonlinear Corrections to Schr\"odinger Dynamics}

From the generalized wave equation,

\begin{equation}
iD_{t}\Psi=-\frac{\zeta}{2m}\triangle\Psi+\left(\frac{1}{4m}\left|\nabla\ln\Psi\right|^{2}+\cdots\right)\Psi
\end{equation}
standard Schr\"odinger dynamics is recovered only when $\zeta=\hbar$,
nonlinear terms lump into the standard quantum potential, and dissipative
degrees of freedom remain dependent only on momenta divergences. Relaxing
these assumptions produces well-defined classes of extensions.

\subsubsection{Nonlinearities of the Doebner--Goldin Type}

Terms such as $\nu\triangle\left|\Psi\right|/\left|\Psi\right|$ and
$i\gamma\nabla\cdot\left(\frac{\nabla\Psi}{\Psi}\right)$ arise naturally
from the non-Hermitian dissipative Hamilton-Jacobi structure. These
resemble the nonlinear Schr\"odinger modifications introduced by Doebner
and Goldin \cite{58}, but here they emerge without ad hoc assumptions
as they are remnants of the underlying dual dynamics.

\subsubsection{Modified Dispersion Relations}

The constant $\zeta$ need not equal $\hbar$. Deviations lead to
modified phase evolution and dispersion relations, analogous to those
explored in generalized quantum frameworks, including non-standard
dispersion and deformed-quantum models \cite{59}.

\subsubsection{Nonlinear Phase-Amplitude Coupling}

Non-stationary imaginary action introduces terms that couple phase
gradients to amplitude gradients. These may become relevant in strongly
confined systems, macroscopic quantum states, or ultrafast/ultracold
regimes. Such couplings resemble families of non-Hermitian hydrodynamic
extensions of quantum mechanics \cite{38}.

\subsection{Modified Quantum Potential and Quantum Geometry}

In equilibrium, the quantum potential takes the familiar form

\begin{equation}
Q=-\frac{\hbar^{2}}{2m}\frac{\triangle\rho^{1/2}}{\rho^{1/2}}
\end{equation}
But away from equilibrium, it becomes

\begin{equation}
Q_{gen}=-\frac{\zeta^{2}}{2m}\left[\frac{\triangle\rho^{1/2}}{\rho^{1/2}}+f\left(\rho,\nabla\rho,\mathcal{S},\nabla\mathcal{S}\right)\right]
\end{equation}
where the correction term $f$ encodes curvature-like contributions
from non-equilibrium complex phase geometry, nonlinear corrections
to the Madelung fluid, and dissipative geometric effects reminiscent
of Weyl-type modifications of quantum mechanics \cite{60}. Thus the
theory naturally leads to geometric generalizations of the quantum
potential and quantum geometry.

\subsection{Objective State Reduction Without Ad Hoc Collapse}

Collapse models such as GRW and CSL introduce external noise fields,
stochastic terms, or tunable collapse rates \cite{54,55}. In contrast,
the present framework predicts collapse dynamically. When system-image-instrument
equilibrium breaks down, the imaginary action develops steep gradients,
and the wavefunction evolves non-unitarily and can localize deterministically.
This produces deterministic, nonlinear collapse in certain regimes,
without stochastic noise and while conserving ensemble probabilities.
Collapse becomes a phase transition in the dual-sector dynamics, not
an externally added rule.

\subsection{Emergent Hilbert Space Structure}

In standard quantum mechanics the Hilbert space is postulated. Here,
it emerges from the dynamics. Complex action gives rise to complex
wavefunction $\Psi=\exp\left(i\mathcal{S}/\zeta\right)$. Dissipative
equilibrium leads to linear PDE for $\Psi$. Balanced dissipation
leads to norm conservation which gives rise to inner product structure.
Physical states lead to equivalence classes of action configurations.
Thus Hilbert space linearity and unitarity break down in regimes with
strong dissipation, rapid phase-gradient evolution, or extremely short
timescales. Related ideas appear in emergent-quantization approaches
\cite{15}, though here the mechanism is explicitly dynamical.

\subsection{Interface With Relativity and Quantum Field Theory}

The framework suggests two natural extensions outlined next.

\subsubsection{Covariant Dissipative HJ Theory}

A four-dimensional complex action $\mathcal{S}\left(x^{\mu}\right)$
with dissipative terms associated with a non-metric tensor could produce
modified relativistic wave equations, complex or PT-symmetric geometric
structures \cite{28} and non-Hermitian covariant extensions of quantum
mechanics.

\subsubsection{Emergent Quantization in Field Theory}

If the complex Hamilton-Jacobi structure generalizes to fields then
second quantization may arise as equilibrium of dissipative field-image
dynamics, canonical commutation relations may emerge as constraints
and vacuum fluctuations may correspond to non-equilibrium excitations
in the image sector. This aligns with broader programs in emergent
quantum field theory .

\section{Conclusion and Outlook}

In this work we developed a unified coupled Hamilton-Jacobi (HJ) framework that forges a structural link between classical dissipative dynamics and quantum mechanics. Starting from the Bateman-Dekker dual-system formulation of loss and gain, we introduced a paired-system (system + environment/image) representation in which every physical degree of freedom is dynamically reflected by the environmental mirror counterpart. When the classical action is extended into the complex domain, the Hamilton-Jacobi equation naturally splits into mutually coupled real (conservative) and imaginary (dissipative) sectors. A central result is that this structure forces a generalized complex wave equation. The standard Schr\"odinger equation then emerges not as a quantization postulate, but as a special equilibrium phase of the dual dissipative dynamics that is obtained precisely when (i) the target system's mass is identical with that of its environmental partner/image, (ii) the coupling potentials are equivalent to momenta divergences and (iii) the environmental image is unconfined (i.e. guiding potential is zero). In this limit, the complex dynamics collapses into linearity, unitarity, and the Hilbert-space structure of ordinary quantum mechanics. This perspective reshapes quantum foundations. Quantum mechanics appears within this framework as a dynamically emergent regime from the equilibrium shadow of a deeper dual dynamics in which (i) unitarity is approximate rather than fundamental, (ii) the quantum potential is a special case of a more general geometric
structure, (iii) nonlinear and non-unitary effects arise naturally,
and (iv) wavefunction collapse becomes a dynamical transition, not
an added axiom. 

Ontologically, the system and its mirror are parts of a single extended entity governed by a complex action. The wavefunction encodes this action; the Born rule corresponds to dissipative equilibration; and measurement corresponds to a transition between dynamical regimes rather than an external intervention. The framework unifies and subsumes multiple previously disconnected approaches (i.e. Bohmian trajectories, PT-symmetric and non-Hermitian systems, stochastic mechanics, nonlinear Schr\"odinger models, and collapse theories) showing them to be limiting
cases or partial descriptions. Importantly, the theory introduces
no ad hoc structures: all features follow from a principled extension
of Hamilton-Jacobi mechanics to systems with dissipation. We also
identified potential empirical signatures, including controlled non-unitarity, modified dispersion relations, and measurable nonlinear amplitude-phase coupling as effects that may be observable in high-coherence platforms, strongly confined systems, or low-temperature experiments. Environmental partitions naturally store the system's past, producing deterministic non-Markovian memory effects. These memory loops yield effective temporal correlations analogous to those invoked in non-Markovian stochastic interpretations of quantum mechanics, providing a dynamical origin for such correlations rather than assuming them as primitives. Future directions include developing a covariant relativistic version of the dissipative HJ formalism, extending the dual framework to quantum fields, exploring generalized quantum potentials, and analyzing non-equilibrium
phases. These may illuminate the behavior of macroscopic quantum systems, decoherence processes, and early-universe dynamics. In summary, the dissipative-complex Hamilton-Jacobi framework provides a coherent and principled extension of quantum mechanics. It reveals quantum theory as one stable dynamical phase in a richer theoretical landscape connecting classical dissipation, complex geometry, and emergent quantum behavior.

The classical–quantum divide in this framework is governed by two factors: (i) the relative strength of the guiding potentials versus the coupling (momenta-divergent) potentials, and (ii) the symmetry properties of the system–image pair, particularly mass symmetry.
Classical behavior emerges whenever the guiding potentials dominate, suppressing the influence of the coupling potentials and effectively erasing dual-sector dynamics. Conversely, quantum behavior arises when the coupling potentials are non-negligible, the coupling is momentum-divergent in form, and the system–image pair exhibits approximate mass symmetry. Under these conditions the dual dynamics stabilize into the symmetric limit where the generalized wave equation reduces to the Schr\"odinger equation. Measurement and wavefunction collapse correspond to abrupt changes in the dominance or functional form of the potentials, driving the system out of the symmetric regime and producing non-unitary evolution.

\bibliographystyle{unsrt}
\bibliography{mybibfile}

\end{document}